\documentclass[12pt]{article}
\usepackage{graphicx}
\usepackage{amssymb}
\usepackage{amsmath}

\newcommand{\bear}{\begin{array}}  
\newcommand {\eear}{\end{array}}
\newcommand{\bea}{\begin{eqnarray}}   
\newcommand{\eea}{\end{eqnarray}}
\newcommand{\beq}{\begin{equation}}   
\newcommand{\eeq}{\end{equation}}
\newcommand{\bef}{\begin{figure}}  \newcommand 
{\eef}{\end{figure}}
\newcommand{\bec}{\begin{center}}  \newcommand 
{\eec}{\end{center}}

\def\lrfp#1#2#3{ \left(\frac{#1}{#2} 
\right)^{#3}}

\def\lrfp#1#2#3{ \left(\frac{#1}{#2} 
\right)^{#3}}

\begin{document}

\begin{titlepage}

\begin{flushright}
IPMU 09-0059 \\
ICRR-Report-544
\end{flushright}

\vskip 1.35cm

\begin{center}
{\large \bf
Non-Gaussianity from Isocurvature Perturbations : \\ Analysis of Trispectrum
 }
\vskip 1.2cm

Etsuko Kawakami$^a$,
Masahiro Kawasaki$^{a,b}$,
Kazunori Nakayama$^a$\\ and
Fuminobu Takahashi$^b$

\vskip 0.4cm

{ \it $^a$Institute for Cosmic Ray Research,
University of Tokyo, Kashiwa 277-8582, Japan}\\
{\it $^b$Institute for the Physics and Mathematics of the Universe,
University of Tokyo, Kashiwa 277-8568, Japan}
\date{\today}

\begin{abstract}
Non-Gaussianity may exist in the CDM isocurvature perturbation.
We provide general expressions for the bispectrum and trispectrum
of both adiabatic and isocurvature pertubations.
We apply our result to the QCD axion case, and found a consistency relation
between the coefficients of the bispectrum and trispectrum :
$\tau_{\rm NL}^{\rm (iso)} \simeq 10^3 [ f_{\rm NL}^{\rm (iso)}]^{4/3}$,
if the axion is dominantly produced by quantum fluctuation.
Thus future observations of the trispectrum, as well as the bispectrum, will be important
for understanding the origin of the CDM and baryon asymmetry.
\end{abstract}


\end{center}
\end{titlepage}

\section{Introduction} \label{intro}

Recent cosmological observations put limits on deviation from the Gaussian cosmological perturbation.
The WMAP collaboration reported a constraint on the non-linearity parameter 
$f_{\rm NL}^{(\rm adi)}$ as
$-9 < f_{\rm NL}^{(\rm adi)} < 111$ at 95\% C.L., while improved analysis by 
another group lead to a slightly tight but consistent result,
$-4 < f_{\rm NL}^{(\rm adi)} < 80$ at 95\% C.L. 
\cite{Smith:2009jr}\footnote{
	In this paper we consider only local type non-Gaussianity.
}.
Thus at present, the observations are consistent with the Gaussian fluctuation.
The standard inflation scenario, where the total density perturbation is created by
the quantum fluctuation of the inflaton,
predicts negligibly small amount of non-Gaussianity,
$f_{\rm NL}^{(\rm adi)} \sim \mathcal O(\epsilon, \eta)$,
where $\epsilon$ and $\eta$ are slow-roll parameters which must be much smaller than
unity for the inflation to last long enough
\cite{Acquaviva:2002ud,Maldacena:2002vr,Seery:2005wm,Yokoyama:2007uu}.
Thus it would be fair to say that
the current observations are consistent with the standard inflation scenario.

However, there are scenarios where non-Gaussianity is enhanced
and can be observed by future experiments.
Let us consider a light scalar field, $\sigma$, which obtains 
a quantum fluctuation during inflation.
In the curvaton scenario~\cite{Mollerach:1989hu,Linde:1996gt,Lyth:2001nq}, 
$\sigma$ is responsible for generation of the density fluctuation.
In this case it is known that non-Gaussianity can be significantly large
depending on its energy density relative to the total energy density of the Universe
at the times of its decay~\cite{Lyth:2002my,Bartolo:2003jx,Lyth:2005fi,Enqvist:2005pg}.
Even in the case where $\sigma$ is not responsible for generating the total adiabatic 
perturbation of the Universe,
the observable level of non-Gaussianity can be imprinted in the 
density perturbation~\cite{Suyama:2008nt}.

Non-Gaussianity may also exist in the isocurvature perturbations
\cite{Linde:1996gt,Bartolo:2001cw,Boubekeur:2005fj}.
This possibility was extensively studied recently for the case of the CDM isocurvature
\cite{Kawasaki:2008sn,Kawasaki:2008pa,Langlois:2008vk} and 
baryonic isocurvature perturbations~\cite{Kawasaki:2008jy,Moroi:2008nn}.
This isocurvature type non-Gaussianity has distinct effects on 
the cosmic microwave background (CMB) bispectrum 
\cite{Kawasaki:2008sn,Kawasaki:2008pa},
and the current constraint reads $f_{\rm NL}^{(\rm iso)}=-5\pm 20$~\cite{Hikage:2008sk}
(definition of $f_{\rm NL}^{(\rm iso)}$ is provided later\footnote{
	Note that our definition of $f_{\rm NL}^{(\rm iso)}$ is same as that used in 
	Refs.~\cite{Kawasaki:2008sn,Kawasaki:2008pa}, but
	differs from that in Ref.~\cite{Hikage:2008sk}.
}).
For instance, non-Gaussianity of isocurvature type can be generated by the axion,
a pseudo Nambu-Goldstone boson associated with the spontaneous breakdown of
Peccei-Quinn (PQ) symmetry, which is introduced in order to solve the strong CP
problem in the quantum chromodynamics
(QCD)~\cite{Peccei:1977hh,Kim:1986ax}.  
The axion is a light scalar field and contributes to the cold dark matter (CDM) of the
Universe~\cite{Preskill:1982cy}. In particular, the axion can have a
large isocurvature perturbation~\cite{Seckel:1985tj,Linde:1991km}.
It was found in Ref.~\cite{Kawasaki:2008sn} that
large enough non-Gaussianity can be generated without conflicting the bound from
power spectrum of the isocurvature perturbations.

In this paper we further investigate non-Gaussianity from isocurvature perturbations,
including not only the bispectrum but also the trispectrum.
So far, the trispectrum of the isocurvature perturbation has not been studied.
As we will see later, however, the trispectrum can become as important as the bispectrum
in certain situations.

This paper is organized as follows.  In Sec.~\ref{formalism}, a
general formalism to calculate bispectrum and trispectrum including isocurvature
perturbations is presented.
In Sec.~\ref{axion} the formalism is applied to the curvaton and axion, and
it is shown that both can induce large non-Gaussianity.
Sec.~\ref{conclusion} is devoted to discussion and conclusions.

\section{Non-Gausianity from adiabatic and isocurvature perturbations} \label{formalism}

\subsection{Non-linear isocurvature perturbations}

We write the perturbed spacetime metric as
\begin{equation}
	ds^2=-{\mathcal N}^2 dt^2 +a^2(t)e^{2\psi} \gamma_{ij} \left ( dx^i + \beta^i dt \right )
	\left ( dx^j + \beta^j dt \right ),
\end{equation}
where ${\mathcal N}$ is the lapse function, $\beta_i$ the shift
vector, $\gamma_{ij}$ the spatial metric, $a(t)$ the background scale
factor, and $\psi$ the curvature perturbation.  On sufficiently large
spatial scales, the curvature perturbation $\psi$ on an arbitrary slicing at
$t=t_f$ is expressed by~\cite{Lyth:2004gb}
\begin{eqnarray}
	\psi (t_f,{\vec x})=N(t_f,t_i:{\vec x})-\log \frac{a(t_f)}{a(t_i)},
	 \label{curvature1}
\end{eqnarray}
where the initial slicing at $t=t_i$ is chosen in such a way that the
curvature perturbations vanish (flat slicing).  Here $N(t_f,t_i:{\vec x})$ is the local $e$-folding number
along the worldline ${\vec x}={\rm const.}$ from $t=t_i$ to $t=t_f$.

We denote by $\zeta$  the curvature perturbation $\psi$ evaluated on the slice where the
total energy density is spatially uniform (uniform-density slicing).
Let us assume that the curvature perturbation is generated by the fluctuation of scalar fields,
$\delta \phi^a$.
From Eq.~(\ref{curvature1}), $\zeta$ can be expanded as
\begin{equation}
	\zeta \simeq N_a \delta \phi^a + \frac{1}{2}N_{ab}\delta \phi^a \delta \phi^b
	+\frac{1}{6}N_{abc} \delta \phi^a \delta \phi^b \delta \phi^c
	+\frac{1}{24}N_{abcd} \delta \phi^a \delta \phi^b \delta \phi^c \delta \phi^d
	+\cdots ,
	\label{deltaN}
\end{equation}
where $N_a \equiv \partial N/\partial \phi^a$ and so on,
and we have included fourth order terms in $\delta \phi$
in order to calculate the trispectrum in a consistent way.

In the similar fashion, we introduce $\zeta_i$ to denote the curvature perturbation on a slice
where $\rho_i$ is uniform ($\delta \rho_i=0$ slicing).
Here $i=c,b,\gamma$ represents the
CDM, baryon and photon, respectively.
If each component does not exchange its energy with the
others, $\zeta_i$ is known to remain constant
for scales larger than the horizon~\cite{Lyth:2004gb}.
We define the (non-linear) isocurvature perturbation between the
$i$-th and $j$-th fluid components as~\cite{Wands:2000dp}
\begin{eqnarray}
	S_{ij} \equiv 3(\zeta_i-\zeta_j). \label{iso1}
\end{eqnarray}
In this paper, we consider the isocurvature perturbation between CDM (baryon) and photon,
$S_{c\gamma} (S_{b\gamma})$.
It is useful to define the effective CDM isocurvature perturbation,
\begin{equation}
	S = S_{c\gamma} + \frac{\Omega_b}{\Omega_c} S_{b\gamma},
\end{equation}
where $\Omega_b/\Omega_{\rm c} \sim 0.2$.
It can be expanded as follows,
\begin{eqnarray}
	S \;\simeq\;S_{a} \delta \phi^a+\frac{1}{2} S_{ab} \delta \phi^a \delta \phi^b
	+\frac{1}{6}S_{abc} \delta \phi^a \delta \phi^b \delta \phi^c
	+\frac{1}{24}S_{abcd} \delta \phi^a \delta \phi^b \delta \phi^c \delta \phi^d
	+\cdots,
	\label{Sm}
\end{eqnarray}
up to the fourth order in $\delta \phi$.
The final curvature perturbation at matter dominated era on sufficiently large scales is 
given by~\cite{Wands:2000dp}
\begin{equation}
	\zeta^{\rm MD} = \zeta +\frac{1}{3}S.   \label{zetaMD}
\end{equation}
Here $\zeta$ is evaluated at deep in the radiation dominated era before the cosmological scales 
enter the horizon, but after the CDM decouples from radiation.\footnote{
	The expression (\ref{zetaMD}) is valid even for the baryonic isocurvature perturbation.
	This is because the baryon number is conserved and the baryons are non-relativistic.
}
Notice that effects of the adiabatic and isocurvature perturbations on 
CMB temperature anisotropy are quite different.
At the Sachs-Wolfe plateau, the CMB 
temperature fluctuation is given by $(\delta T/T) = (1/5)\zeta +(2/5)S$,
from which we can see that relative importance of the isocurvature perturbation is enhanced
with respect to its effect on the density perturbation. 

For simplicity, we assume that masses of $\{\phi_a\}$ are
light enough, and that the fluctuations are Gaussian and independent to each other. 
Then the correlation functions are given by the following form,
\begin{equation}
	\langle \delta \phi^a_{\vec k_1} \delta \phi^b_{\vec k_2} 
	\rangle\;=\;(2\pi)^3\,P_{\delta \phi}(k_1)
	\delta(\vec k_1+\vec k_2)\delta^{ab}
	\label{phiphi}
\end{equation}
with
\beq
	P_{\delta \phi}(k) \;\simeq\; \frac{H_{\rm inf}^2}{2k^3},
\eeq
where $k$ denotes a comoving wavenumber, and 
$H_{\rm inf}$ is the Hubble parameter during inflation.  For
later use, we also define the following:
\beq
	\Delta_{\delta \phi}^2 \;\equiv\; \frac{k^3}{2\pi^2} P_{\delta \phi}(k) \simeq \lrfp{H_{\rm inf}}{2\pi}{2}.
\eeq
Since $\delta \phi^a$ are Gaussian variables, their three point functions vanish~:
\begin{equation}
	\langle \delta \phi^a_{\vec k_1} \delta \phi^b_{\vec k_2} \delta \phi^c_{\vec k_3} \rangle= 0.
\end{equation}

Using the expansion (\ref{deltaN}) and (\ref{Sm}), we can calculate the power spectrum,
bispectrum and trispectrum of the curvature perturbation (\ref{zetaMD}).
In the following we derive general expressions for the bispectrum and trispectrum
including both adiabatic and isocurvature perturbations.

\subsection{Power spectrum}

First we calculate the power spectrum of the curvature perturbation.
We define the followings,
\begin{gather}
	\langle \zeta_{\vec k_1} \zeta_{\vec k_2} \rangle \equiv
	(2\pi)^3 \delta(\vec k_1+\vec k_2) P_\zeta(k_1), \\
	\langle \zeta_{\vec k_1} S_{\vec k_2} \rangle \equiv
	(2\pi)^3 \delta(\vec k_1+\vec k_2) P_{\zeta S}(k_1), \\
	\langle S_{\vec k_1} S_{\vec k_2} \rangle \equiv
	(2\pi)^3 \delta(\vec k_1+\vec k_2) P_{S}(k_1).
\end{gather}
Using (\ref{deltaN}) and (\ref{Sm}), we obtain
\begin{gather}
	P_\zeta(k)=N_aN_a P_{\delta \phi}(k) + \frac{1}{2}N_{ab}N_{ab}
	\int \frac{d^3 \vec k'}{(2\pi)^3}P_{\delta \phi}(k')P_{\delta \phi}(|\vec k-\vec k'|),\\
	P_{\zeta S}(k)=N_aS_a P_{\delta \phi}(k) + \frac{1}{2}N_{ab}S_{ab}
	\int \frac{d^3 \vec k'}{(2\pi)^3}P_{\delta \phi}(k')P_{\delta \phi}(|\vec k-\vec k'|),\\
	P_S(k)=S_aS_a P_{\delta \phi}(k) + \frac{1}{2}S_{ab}S_{ab}
	\int \frac{d^3 \vec k'}{(2\pi)^3}P_{\delta \phi}(k')P_{\delta \phi}(|\vec k-\vec k'|).	
\end{gather}
After performing the integration, the spectra $P_\zeta$, $P_{\zeta S}$ and $P_S$ can be expressed as
\begin{eqnarray}
	P_\zeta(k) \simeq [ N_a N_a + N_{ab}N_{ab}\Delta_{\delta \phi}^2 \ln (kL) ]
	P_{\delta \phi}(k), \\
	P_{\zeta S}(k) \simeq [ N_a S_a + N_{ab}S_{ab}\Delta_{\delta \phi}^2 \ln (kL) ] 
	P_{\delta \phi}(k), \\
	P_{S}(k) \simeq [ S_a S_a + S_{ab}S_{ab}\Delta_{\delta \phi}^2 \ln (kL) ] P_{\delta \phi}(k).
\end{eqnarray}
Here we have introduced an infrared cutoff $L$, which is taken to be of the order of the present Hubble
horizon scale~\cite{Lyth:1991ub,Lyth:2007jh}.
The recent WMAP5 result shows that the isocurvature perturbations must be subdominant component
and the 95\% C.L. limits read $P_S(k_*)/P_\zeta(k_*) \lesssim 0.19$ for the uncorrelated case,
and $P_S(k_*)/P_\zeta(k_*) \lesssim 0.011$ for the totally anti-correlated case,
where $k_* = 0.002~{\rm Mpc}^{-1}$ is the pivot scale~\cite{Komatsu:2008hk}.
Thus we can safely take $P_\zeta (k) \simeq N_aN_a P_{\delta \phi}(k)$, which is subject to
the WMAP normalization condition, $\Delta_\zeta^2(k_*)
\equiv (k_*^3/2\pi^2)P_\zeta(k_*)\simeq 2.5\times 10^{-9}$.

\subsection{Bispectrum}

We define the bispectrum of the curvature perturbation $B_\zeta^{\rm MD}$ as
\begin{eqnarray}
	\langle \zeta_{ {\vec k_1}}^{\rm MD} \zeta_{ {\vec k_2}}^{\rm MD} 
	\zeta_{ {\vec k_3}}^{\rm MD} \rangle
	\equiv (2\pi)^3\delta(\vec k_1+\vec k_2+\vec k_3)~B^{\rm MD}_\zeta (k_1,k_2,k_3).
\end{eqnarray}
Notice that this contains four kind of terms taking account of the expression (\ref{zetaMD}),
such as $\langle \zeta\zeta\zeta \rangle, \langle \zeta\zeta S \rangle,
\langle \zeta SS \rangle$ and $\langle SSS \rangle$.
Let us therefore divide $B_\zeta^{\rm MD}$ into four parts depending on 
the origin of each term as 
\begin{equation}
	B^{\rm MD}_\zeta (k_1,k_2,k_3) = B_{\zeta\zeta\zeta}+B_{\zeta\zeta S}+
	B_{\zeta SS}+B_{SSS}.  \label{Bzeta}
\end{equation}
Full expression for each term is a bit complicated, and given in the Appendix.
We define non-linerity parameters as~\cite{Kawasaki:2008pa}
\begin{equation}
\begin{split}
	B_{\zeta \zeta \zeta} &= \frac{6}{5}f_{\rm NL}^{(\rm adi)}
	[P_\zeta(k_1)P_\zeta(k_2)+{\rm 2~perms}], \\
	B_{\zeta \zeta S} &= \frac{6}{5}f_{\rm NL}^{(\rm cor1)}
	[P_\zeta(k_1)P_\zeta(k_2)+{\rm 2~perms}], \\
	B_{\zeta SS} &= \frac{6}{5}f_{\rm NL}^{(\rm cor2)}
	[P_\zeta(k_1)P_\zeta(k_2)+{\rm 2~perms}], \\
	B_{SSS} &= \frac{6}{5}f_{\rm NL}^{(\rm iso)}
	[P_\zeta(k_1)P_\zeta(k_2)+{\rm 2~perms}], 
\end{split}
\end{equation}
and using the results given in the Appendix, we obtain
\begin{eqnarray}
	&&\frac{6}{5}f_{\rm NL}^{(\rm adi)} = \frac{1}{(N_aN_a)^2}
	\left[ N_a N_bN_{ab} + (N_{ab}N_{bc}N_{ca}+2N_aN_{bc}N_{abc})
		\Delta_{\delta \phi}^2 \ln (k_bL) \right], \label{fNLadi} \\
	&&\frac{6}{5}f_{\rm NL}^{(\rm cor1)} = \frac{1}{3(N_aN_a)^2}
		\left[N_a N_bS_{ab} + 2N_{ab}N_a S_b +\{3N_{ab}N_{bc}S_{ca} \right. \nonumber \\
		&&~~~~~~~~~~+2(S_{a}N_{bc}N_{abc}+N_{a}S_{bc}N_{abc}+N_{a}N_{bc}S_{abc} ) \}
		\Delta_{\delta \phi}^2 \ln (k_bL)\left. \right],  \label{fNLcor1} \\
	&&\frac{6}{5}f_{\rm NL}^{(\rm cor2)} = \frac{1}{9(N_aN_a)^2}
		\left[ S_a S_bN_{ab} + 2S_{ab}S_a N_b +\{3S_{ab}S_{bc}N_{ca} \right. \nonumber \\
		&&~~~~~~~~~~+2(N_{a}S_{bc}S_{abc}+S_{a}N_{bc}S_{abc}+S_{a}S_{bc}N_{abc} ) \}
		\Delta_{\delta \phi}^2 \ln (k_bL)\left. \right],  \label{fNLcor2} \\
	&&\frac{6}{5}f_{\rm NL}^{(\rm iso)} = \frac{1}{27(N_aN_a)^2}
	\left[ S_a S_bS_{ab} + (S_{ab}S_{bc}S_{ca}+2S_aS_{bc}S_{abc})
		\Delta_{\delta \phi}^2 \ln (k_bL) \right],  \label{fNLiso}
\end{eqnarray}
where $k_b={\rm min}\{ k_i\}~(i=1,2,3)$.
The first term in $f_{\rm NL}^{(\rm adi)}$ was obtained in Ref.~\cite{Lyth:2005fi}.
Here we have included terms containing the third derivatives of $N$ and $S$.
In a simple case where the scalar potentials 
are given by the quadratic form, such terms including the third derivatives can be neglected.
In the case of the standard single field slow-roll inflation, Eq.~(\ref{fNLadi}) gives 
$f_{\rm NL}\sim \mathcal O(\epsilon,\eta)$ where $\epsilon$ and $\eta$ are slow-roll parameters, 
as explicitly shown in Ref.~\cite{Byrnes:2006vq,Yokoyama:2007uu},
which coincides with earlier investigations~\cite{Maldacena:2002vr}.
Expressions (\ref{fNLadi})-(\ref{fNLiso}) generalize the results of 
Refs.~\cite{Kawasaki:2008pa,Langlois:2008vk}.

\subsection{Trispectrum}

Similarly to the case of bispectrum, the trispectrum of the curvature perturbation
$T^{\rm MD}_\zeta$ is defined as
\begin{eqnarray}
	\langle \zeta_{ {\vec k_1}}^{\rm MD} \zeta_{ {\vec k_2}}^{\rm MD} 
	\zeta_{ {\vec k_3}}^{\rm MD} \zeta_{ {\vec k_4}}^{\rm MD} \rangle
	\equiv (2\pi)^3\delta(\vec k_1+\vec k_2+\vec k_3+\vec k_4)
	T^{\rm MD}_\zeta (k_1,k_2,k_3,k_4).
\end{eqnarray}
This contains five kinds of terms, such as 
$\langle \zeta\zeta\zeta\zeta \rangle, \langle \zeta\zeta\zeta S \rangle, \langle \zeta\zeta SS\rangle,
\langle \zeta SSS \rangle$ and $\langle SSSS \rangle$, and hence can be divided as
\begin{equation}
	T^{\rm MD}_\zeta (k_1,k_2,k_3,k_4) = T_{\zeta\zeta\zeta\zeta}+
	T_{\zeta\zeta\zeta S}+T_{\zeta\zeta SS}+T_{\zeta SSS}+T_{SSSS}. \label{Tzeta}
\end{equation}
Full expressions for these terms are given in the Appendix.
Defining non-linerity parameters $\tau_{\rm NL}$ and $g_{\rm NL}$,
which characterize the trispectrum including isocurvature perturbations as
\begin{equation}
\begin{split}
	T_{\zeta \zeta \zeta \zeta} &=\tau_{\rm NL}^{(\rm adi)}
	[P_\zeta(k_{13})P_\zeta(k_3)P_\zeta(k_4)+{\rm 11~perms}]\\
	&~~~+\frac{54}{25}g_{\rm NL}^{(\rm adi)}
	[P_\zeta(k_1)P_\zeta(k_2)P_\zeta(k_3)+{\rm 3~perms}] , \\
\end{split}
\end{equation}
\begin{equation}
\begin{split}
	T_{\zeta \zeta\zeta S} &= \tau_{\rm NL}^{(\rm cor1)}
	[P_\zeta(k_{13})P_\zeta(k_3)P_\zeta(k_4)+{\rm 11~perms}] \\
	&~~~+\frac{54}{25}g_{\rm NL}^{(\rm cor1)}
	[P_\zeta(k_1)P_\zeta(k_2)P_\zeta(k_3)+{\rm 3~perms}], \\
\end{split}
\end{equation}
\begin{equation}
\begin{split}
	T_{\zeta\zeta SS} &=\tau_{\rm NL}^{(\rm cor2)}
	[P_\zeta(k_{13})P_\zeta(k_3)P_\zeta(k_4)+{\rm 11~perms}] \\
	&~~~+\frac{54}{25}g_{\rm NL}^{(\rm cor2)}
	[P_\zeta(k_1)P_\zeta(k_2)P_\zeta(k_3)+{\rm 3~perms}], \\
\end{split}
\end{equation}
\begin{equation}
\begin{split}
	T_{\zeta SSS} &=\tau_{\rm NL}^{(\rm cor3)}
	[P_\zeta(k_{13})P_\zeta(k_3)P_\zeta(k_4)+{\rm 11~perms}]\\
	&~~~+\frac{54}{25}g_{\rm NL}^{(\rm cor3)}
	[P_\zeta(k_1)P_\zeta(k_2)P_\zeta(k_3)+{\rm 3~perms}], \\
\end{split}
\end{equation}
\begin{equation}
\begin{split}
	T_{SSSS} &=\tau_{\rm NL}^{(\rm iso)}
	[P_\zeta(k_{13})P_\zeta(k_3)P_\zeta(k_4)+{\rm 11~perms}] \\
	&~~~+\frac{54}{25}g_{\rm NL}^{(\rm iso)}
	[P_\zeta(k_1)P_\zeta(k_2)P_\zeta(k_3)+{\rm 3~perms}],
\end{split}
\end{equation}
where we have used the notation $k_{ij}\equiv |\vec k_i +\vec k_j|$.
Combining the above expressions with the results given in the Appendix, we obtain
\begin{eqnarray}
&&\tau_{\rm NL}^{(\rm adi)} = \frac{1}{(N_aN_a)^3}
	\left[ N_a N_bN_{ac}N_{bc} 
	+N_{ab}N_{bc}N_{cd}N_{da} \Delta_{\delta \phi}^2 \ln (k_tL) \right. \nonumber \\ 
	&&\left.+2N_aN_{bd}N_{cd}N_{abc}\Delta_{\delta \phi}^2 \ln (k_bL) + 
	\left( 2N_aN_{ab}N_{cd}N_{bcd}+N_aN_bN_{acd}N_{bcd} \right)\Delta_{\delta \phi}^2 \ln (kL)
	\right], 
\end{eqnarray}
\begin{eqnarray}
&&\tau_{\rm NL}^{(\rm cor1)} = \frac{1}{3(N_aN_a)^3}
	\left[ 2(N_a S_bN_{ac}N_{bc} + N_a N_bN_{ac}S_{bc})
	+4N_{ab}N_{bc}N_{cd}S_{da} \Delta_{\delta \phi}^2 \ln (k_tL) \right. \nonumber \\ 
	&&~~~~~+2 (S_{abc}N_{ad}N_{cd}N_{b}+2N_{abc}S_{ad}N_{cd}N_{b}
					+N_{abc}N_{ad}N_{cd}S_{b})\Delta_{\delta \phi}^2 \ln (k_bL) \nonumber \\
	&&~~~~+\left \{2(S_{abc}N_{ab}N_{cd}N_{d}+N_{abc}S_{ab}N_{cd}N_{d}
					+N_{abc}N_{ab}S_{cd}N_{d}+N_{abc}N_{ab}N_{cd}S_{d}  )\right. \nonumber \\
	&&~~~~+\left. \left. (2N_{abc}S_{abd}N_{c}N_{d}+2N_{abc}N_{abd}N_{c}S_{d})
	 \right \} \Delta_{\delta \phi}^2 \ln (kL)
	\right], 
\end{eqnarray}
\begin{eqnarray}
&&\tau_{\rm NL}^{(\rm cor2)} = \frac{1}{9(N_aN_a)^3}
	\left[ (N_aN_bS_{ac}S_{bc}+2N_aN_{ab}S_{bc}S_c
		+2N_aS_{ab}N_{bc}S_c+S_aS_bN_{ac}N_{bc})
		\right. \nonumber \\
	&&~~~~~+(4N_{ab}N_{bc}S_{cd}S_{da} +2N_{ab}S_{bc}N_{cd}S_{da}  )
		\Delta_{\delta \phi}^2 \ln (k_tL) \nonumber \\ 
	&&~~~~+2 (2N_{abc}N_{ad}S_{cd}S_{b}+N_{abc}S_{ad}S_{cd}N_{b}
		+S_{abc}N_{ad}N_{cd}S_{b}+2S_{abc}N_{ab}S_{cd}N_{d})\Delta_{\delta \phi}^2 \ln (k_bL) 
		\nonumber \\
	&&~~~~~+\left \{ 2(S_{abc}S_{ab}N_{cd}N_{d}+S_{abc}N_{ab}N_{cd}S_{d}
		+S_{abc}N_{ab}S_{cd}N_{d}+N_{abc}S_{ab}N_{cd}S_{d}\right. \\
	&&~~~~~~~~~+N_{abc}S_{ab}S_{cd}N_{d}+N_{abc}N_{ab}S_{cd}S_{d}  )\nonumber \\
	&&~~~~+\left. \left. (S_{abc}S_{abd}N_{c}N_{d}+ 2N_{abc}S_{abd}N_{c}S_{d}
		+2N_{abc}S_{abd}S_{c}N_{d}+N_{abc}N_{abd}S_{c}S_{d}  ) 
	 \right \}\Delta_{\delta \phi}^2 \ln (kL)
	\right], \nonumber
\end{eqnarray}
\begin{eqnarray}
&&\tau_{\rm NL}^{(\rm cor3)} = \frac{1}{27(N_aN_a)^3}
	\left[ 2(S_a N_bS_{ac}S_{bc} + S_a S_bS_{ac}N_{bc})
	+4S_{ab}S_{bc}S_{cd}N_{da} \Delta_{\delta \phi}^2 \ln (k_tL) \right. \nonumber \\ 
	&&~~~~~+2 (N_{abc}S_{ad}S_{cd}S_{b}+2S_{abc}N_{ad}S_{cd}S_{b}
					+S_{abc}S_{ad}S_{cd}N_{b})\Delta_{\delta \phi}^2 \ln (k_bL) \nonumber \\
	&&~~~~+\left \{ 2(N_{abc}S_{ab}S_{cd}S_{d}+S_{abc}N_{ab}S_{cd}S_{d}
					+S_{abc}S_{ab}N_{cd}S_{d}+S_{abc}S_{ab}S_{cd}N_{d} )\right.  \nonumber \\
	&&~~~~+\left. \left. (2S_{abc}N_{abd}S_{c}S_{d}+2S_{abc}S_{abd}S_{c}N_{d})
	 \right \} \Delta_{\delta \phi}^2 \ln (kL)
	\right], 
\end{eqnarray}
\begin{eqnarray}
&&\tau_{\rm NL}^{(\rm iso)} = \frac{1}{81(N_aN_a)^3}
	\left[ S_a S_bS_{ac}S_{bc} 
	+S_{ab}S_{bc}S_{cd}S_{da} \Delta_{\delta \phi}^2 \ln (k_tL) \right. \nonumber \\ 
	&&\left.+2S_aS_{bd}S_{cd}S_{abc}\Delta_{\delta \phi}^2 \ln (k_bL) + 
	\left(2S_aS_{ab}S_{cd}S_{bcd}+S_aS_bS_{acd}S_{bcd} \right)\Delta_{\delta \phi}^2 \ln (kL)
	\right],
\end{eqnarray}
for $\tau_{\rm NL}$, where $k_t ={\rm min}~\{ k_{ij},k_l \}~(i,j,l=1,2,3,4)$, 
 where $k_b^3 ={\rm min}\{ k_i k_j^2 \}$ and
$k^3 = {\rm min}\{ k_i k_j k_l \}$, and
\begin{eqnarray}
&&\frac{54}{25}g_{\rm NL}^{(\rm adi)} = \frac{1}{(N_aN_a)^3}
	\left[ N_a N_bN_{c}N_{abc} 
	+3N_{a}N_{bd}N_{cd}N_{abc} \Delta_{\delta \phi}^2 \ln (k_bL) \right. \nonumber \\ 
	&&~~~~~\left.+3N_aN_{b}N_{cd}N_{abcd}\Delta_{\delta \phi}^2 \ln (kL)\right], \\
&&\frac{54}{25}g_{\rm NL}^{(\rm cor1)} = \frac{1}{3(N_aN_a)^3}
	\left[ 3N_{a}N_{b}S_{c}N_{abc}+S_{abc}N_aN_bN_c \right.\nonumber \\
	&&~~~~~+3(S_{abc}N_{ad}N_{cd}N_{b}+2N_{abc}S_{ad}N_{cd}N_{b}+N_{abc}N_{ad}N_{cd}S_{b})
		\left. \Delta_{\delta \phi}^2 \ln (k_bL) \right. \nonumber \\ 
	&&~~~~~\left.+3(S_{abcd}N_{ab}N_{c}N_{d}+N_{abcd}S_{ab}N_{c}N_{d}
	   	+2N_{abcd}N_{ab}N_{c}S_{d}) \Delta_{\delta \phi}^2 \ln (kL)\right], \\
&&\frac{54}{25}g_{\rm NL}^{(\rm cor2)} = \frac{1}{9(N_aN_a)^3}
	\left[ 3N_{a}S_{b}S_{c}N_{abc}+3S_{abc}S_aN_bN_c
	\right. \\
	&&~~~~~+3(2N_{abc}N_{ad}S_{cd}S_{b}+N_{abc}S_{ad}S_{cd}N_{b}
		+S_{abc}N_{ad}N_{cd}S_{b}+2S_{abc}N_{ab}S_{cd}N_{d}) 
		\left. \Delta_{\delta \phi}^2 \ln (k_bL) \right. \nonumber \\ 
	&&~~~\left.+3(N_{abcd}N_{ab}S_{c}S_{d}+2N_{abcd}S_{ab}N_{c}S_{d}
	      +2S_{abcd}N_{ab}N_{c}S_{d}+S_{abcd}S_{ab}N_{c}N_{d}) \Delta_{\delta \phi}^2 \ln (kL)\right], 
	      \nonumber \\
&&\frac{54}{25}g_{\rm NL}^{(\rm cor3)} = \frac{1}{27(N_aN_a)^3}
	\left[ 3S_{a}S_{b}N_{c}S_{abc}+N_{abc}S_aS_bS_c \right.\nonumber \\
	&&~~~~~+3(N_{abc}S_{ad}S_{cd}S_{b}+2S_{abc}N_{ad}S_{cd}S_{b}+S_{abc}S_{ad}S_{cd}N_{b})
		\left. \Delta_{\delta \phi}^2 \ln (k_bL) \right. \nonumber \\ 
	&&~~~~~\left.+3(N_{abcd}S_{ab}S_{c}S_{d}+S_{abcd}N_{ab}S_{c}S_{d}
	   	+2S_{abcd}S_{ab}S_{c}N_{d}) \Delta_{\delta \phi}^2 \ln (kL)\right], \\
&&\frac{54}{25}g_{\rm NL}^{(\rm iso)} = \frac{1}{81(N_aN_a)^3}
	\left[ S_a S_bS_{c}S_{abc} 
	+3S_{a}S_{bd}S_{cd}S_{abc} \Delta_{\delta \phi}^2 \ln (k_bL) \right. \nonumber \\ 
	&&~~~~~\left.+3S_aS_{b}S_{cd}S_{abcd}\Delta_{\delta \phi}^2 \ln (kL)\right], 
\end{eqnarray}
for $g_{\rm NL}$.
The first terms in $\tau_{\rm NL}^{\rm (adi)}$ and $g_{\rm NL}^{\rm (adi)}$ 
were given in Ref.~\cite{Byrnes:2006vq}, and next order expressions were
presented in Ref.~\cite{Suyama:2008nt}.
In the case of quadratic scalar potentials for $\phi^a$, $g_{\rm NL}$ are 
suppressed compared with $\tau_{\rm NL}$
since $g_{\rm NL}$ always include third order derivatives such as $N_{abc}$ and $S_{abc}$.
Note that these $\tau_{\rm NL}$ and $g_{\rm NL}$ are defined
through the curvature perturbation in the matter dominated era at large scales.
As already mentioned, the effect on the temperature fluctuation is enhanced at the Sachs-Wolfe plateau
in the case of the isocurvature perturbation, compared with the adiabatic curvature perturbation.
For example, at large scales, the primordial non-Gaussianity with $\tau_{\rm NL}^{\rm (iso)}=1$
may have about 1000 times enhanced effect
on the trispectrum of the temperature fluctuation on large scale,
than that with $\tau_{\rm NL}^{\rm (adi)}=1$.
In order to see how they exhibit themselves on the CMB trispectrum beyond the Sachs-Wolfe plateau, 
we must connect the curvature and isocurvature perturbations to the temperature fluctuation
using the transfer functions,
as we did in the case of bispectrum~\cite{Kawasaki:2008sn,Kawasaki:2008pa}.
Thus their effects on the CMB temperature anisotropy may be quite characteristic.
This will be studied in a separate paper.

\section{Application} \label{axion}


We have provided general formulae of
$f_{\rm NL}$, $\tau_{\rm NL}$ and $g_{\rm NL}$
in the general multi-field case in order to keep our arguments most generic.
Although these formulae are a bit complicated, we see that in this section they
can be reduced to a simple form in some models.

Let us consider a case where both the inflaton $\phi$ and another light scalar $\sigma$ 
obtain quantum fluctuations during inflation.
We assume that $\sigma$ has a quadratic potential, and 
that the inflaton generated negligible amount of non-Gaussianity. 
Then terms involving $N_{\phi \phi}$, $N_{\sigma \sigma \sigma}$ and higher
derivatives can be neglected.
If the $\sigma$ field decays into radiation at some epoch well before the big-bang nucleosynthesis,
the primordial fluctuation of $\sigma$ is inherited in the adiabatic perturbation,
and so is its non-Gaussianity.
Then non-linearity parameters of the adiabatic type are given by
\begin{equation}
	\frac{6}{5}f_{\rm NL}^{\rm (adi)}=
	\frac{N_\sigma^2N_{\sigma\sigma}+N_{\sigma\sigma}^3\Delta_{\delta \phi}^2\ln(k_bL)}
	{(N_\phi^2 +N_\sigma^2)^2},  \label{fNLsimple}
\end{equation}
and
\begin{equation}
	\tau_{\rm NL}^{\rm (adi)}=
	\frac{N_\sigma^2N_{\sigma\sigma}^2+N_{\sigma\sigma}^4\Delta_{\delta \phi}^2\ln(k_tL)}
	{(N_\phi^2+N_\sigma^2)^3}.
\end{equation}
The other non-linearity parameter, $g_{\rm NL}$, contains higher derivative terms and hence
the dominant contribution to the trispectrum comes from $\tau_{\rm NL}$.
We study this case in the next subsection.

Let us consider another case where $\sigma$ is stable and contributes to the CDM at present epoch.
Again the potential of $\sigma$ is assumed to be quadratic in $\sigma$.
In this case, the fluctuation of $\sigma$ becomes the CDM isocurvature perturbation.
Notice that the energy density of $\sigma$ is completely negligible
when cosmological scales of interest enter the horizon in the radiation dominated era, and hence
$N_{\sigma}, N_{\sigma\sigma}$ and so on, can be safely neglected.
Therefore this case results in the non-Gaussianity of isocurvature type,
\begin{equation}
	\frac{6}{5}f_{\rm NL}^{\rm (iso)}=
	\frac{S_\sigma^2S_{\sigma\sigma}+S_{\sigma\sigma}^3\Delta_{\delta \phi}^2\ln(k_bL)}
	{27N_\phi^4},  \label{fNLsimple}
\end{equation}
and
\begin{equation}
	\tau_{\rm NL}^{\rm (iso)}=
	\frac{S_\sigma^2S_{\sigma\sigma}^2+S_{\sigma\sigma}^4\Delta_{\delta \phi}^2\ln(k_tL)}
	{81N_\phi^6}.  \label{tauNLsimple}
\end{equation}
As we will see below, $\sigma$ can be identified as the axion.
In general, calculating the coefficients $S_\sigma$, $S_{\sigma \sigma}$, and so on
may be a bothersome task.
In simple cases, however, we can solve the evolution of $\sigma$ analytically
(see Refs.~\cite{Kawasaki:2008sn,Kawasaki:2008jy,Kawasaki:2008pa} for concrete examples).
For instance, for the case of quadratic potential, we obtain 
$S_\sigma \sim 1/\sigma$ and $S_{\sigma\sigma} \sim 1/\sigma^2$ (cf. Eq.(\ref{SCDM})).

\subsection{Curvaton}

Now let us identify $\sigma$ with a curvaton, which dominantly contributes to the
final curvature perturbation.
We assume that some fraction of the DM (denoted by $X$) is produced by the inflaton decay, 
and the remaining DM (denoted by $Y$) originates from thermal bath after the curvaton decays.
Then $X$ has correlated isocurvature perturbation with the adiabatic one,
since it fluctuates in the same way as inflaton, not as the curvaton (or equivalently, radiation).
In this case the fluctuation of $\sigma$ leads to both 
adiabatic and isocurvature type non-Gaussianities.

For the bispectrum, dominant contributions are $f_{\rm NL}^{\rm (adi)}$
and $f_{\rm NL}^{\rm (cor1)}$, given as~\cite{Bartolo:2003jx,Lyth:2005fi,Kawasaki:2008pa}
\begin{equation}
	\frac{6}{5}f_{\rm NL}^{\rm (adi)}=\frac{1}{2R}\left( 3-4R-2R^2\right),
\end{equation}
\begin{equation}
	\frac{6}{5}f_{\rm NL}^{\rm (cor1)}=-\frac{3\epsilon}{2R}
	\left[ 3-4R-2R^2 -2R(1-\epsilon) \right]
\end{equation}
where $R$ roughly denotes the energy fraction of $\sigma$ field at its decay epoch :
$R=3\rho_\sigma/(4\rho_r + 3\rho_\sigma)$ with $\rho_r$ $(\rho_\sigma)$
denoting the energy density of the radiation (curvaton) at the epoch of curvaton decay, and
$\epsilon$ denotes the fraction of $X$ to the total DM abundance.
From WMAP5 constraint on the isocurvature fraction, it is limited as
$\epsilon \lesssim 0.035$.
Here we have assumed that classical displacement of the curvaton field 
is much larger than the amplitude of the quantum fluctuation, 
and hence neglected second term in Eq.~(\ref{fNLsimple}).
Although $f_{\rm NL}^{\rm (cor1)}$ is suppressed by small coefficient $\epsilon$,
it can have visible effect on the temperature fluctuation because of the 
enhancement of the isocurvature perturbation 
at large scale~\cite{Kawasaki:2008sn,Kawasaki:2008pa}.

Similarly, we can calculate $\tau_{\rm NL}^{\rm (adi)}$ and $\tau_{\rm NL}^{\rm (cor1)}$.
The result is
\begin{equation}
	\tau_{\rm NL}^{\rm (adi)}=\frac{36}{25}\left[  f_{\rm NL}^{\rm (adi)} \right]^2
	= \frac{1}{4R^2}\left( 3-4R-2R^2\right)^2,
\end{equation}
for the adiabatic type as in the usual case.
On the other hand, we also obtain
\begin{equation}
	\tau_{\rm NL}^{\rm (cor1)}=-\frac{\epsilon}{R^2}( 3-4R-2R^2 )
	\left[  3-4R-2R^2-3R(1-\epsilon) \right].
\end{equation}
Other types of $\tau_{\rm NL}$'s ($\tau_{\rm NL}^{\rm (cor2)},\tau_{\rm NL}^{\rm (cor3)},
\tau_{\rm NL}^{\rm (iso)}$) contain more powers of $\epsilon$,
and hence have only subdominant effect on the trispectrum.
In the small $R$ limit, we find another consistency relation 
$\tau_{\rm NL}^{\rm (cor1)}=(48/25)f_{\rm NL}^{\rm (adi)}f_{\rm NL}^{\rm (cor1)}$,
which, if confirmed, will support the curvaton scenario.

\subsection{Axion}

In this section we apply our results to the axion.
The axion, $a$, is a pseudo Nambu-Goldstone boson associated with the spontaneous breaking of
the PQ symmetry. The breaking scale of the PQ symmetry, $F_a$,
is constrained from various experiments as well as astrophysical and cosmological considerations.
From the observation of SN1987A, $F_a\gtrsim 10^{10}~$GeV is required~\cite{Raffelt:1990yz}.
On the other hand, the upper bound is provided by the cosmological argument.
Since the axion obtains a tiny mass after the QCD phase transition,
it begins to oscillate coherently since then and it behaves as non-relativistic matter.
The lifetime of the axion is much longer than the present age of the Universe,
and hence its coherent oscillations contribute to the present dark matter abundance.
The abundance of the axion relative to the dark matter abundance,
$r \equiv \Omega_a/\Omega_{c}$, is estimated as \cite{Turner:1985si}
\begin{equation}
	r \;\simeq\; 1.8 \times 10^{-2}\left ( \frac{F_a}{10^{12}~{\rm GeV}} \right )^{-0.82}
	\left ( \frac{a_*}{10^{11}~{\rm GeV}} \right )^2
	\left ( \frac{0.11}{\Omega_{c}h^2} \right ),
\end{equation}
where we have defined
\begin{equation}
	a_* \;\equiv\; {\rm max} \left \{ F_a \theta,~~\frac{H_{\rm inf}}{2\pi} \right \}.  \label{a*}
\end{equation}
Here $\theta$ denotes the initial misalignment angle of the axion.
Thus the PQ scale is constrained as $F_a \lesssim \theta^{-1.7}10^{12}$~GeV.

\begin{figure}[h]
 \begin{center}
   \includegraphics[width=0.7\linewidth]{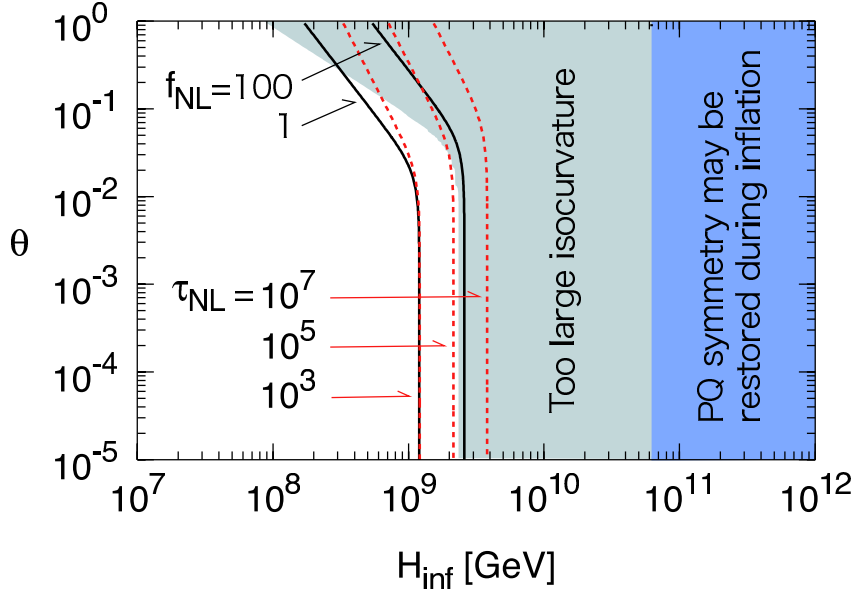} 
   \caption{
  	Contours of $f_{\rm NL}^{(\rm iso)}=1$ and $100$ for $F_a = 10^{10}~$GeV (black solid lines)
	and $\tau_{\rm NL}^{(\rm iso)}=10^3,10^5$ and $10^7$ (red dashed lines).
	The gray shaded region is excluded from isocurvature constraint.
	In the blue region the PQ symmetry may be restored during inflation, and if this is the case,
	neither isocurvature fluctuation nor non-Gaussianity will arise.
   }
   \label{fig:NG-Fa10}
 \end{center}
\end{figure}

\begin{figure}[h]
 \begin{center}
   \includegraphics[width=0.7\linewidth]{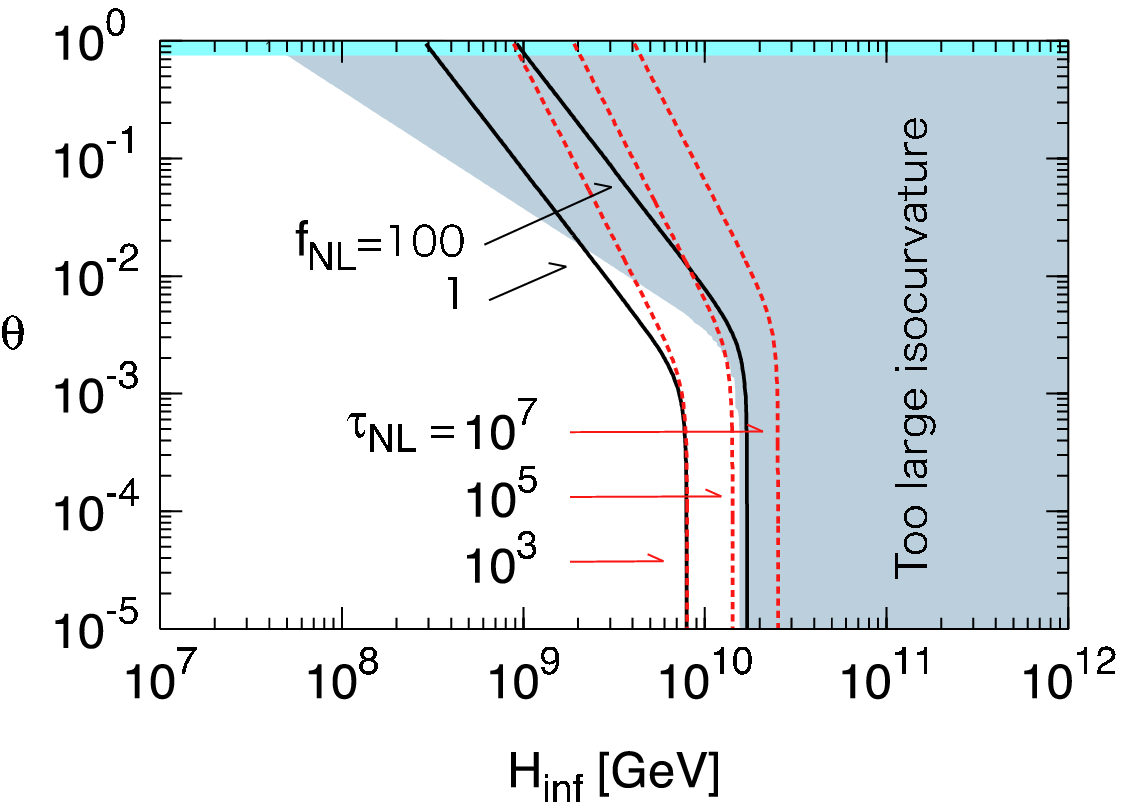} 
   \caption{
  	 Same as Fig.~\ref{fig:NG-Fa10}, but for $F_a = 10^{12}~$GeV.
	 The upper blue region ($\theta \gtrsim 0.7$) is excluded from the axion overproduction.
   }
   \label{fig:NG-Fa12}
 \end{center}
\end{figure}

If the PQ symmetry is already broken before or during inflation and if it is never restored after inflation,
the axion obtains quantum fluctuations during inflation and
its fluctuation becomes the CDM isocurvature fluctuation.
This is realized if the inflation scale and the subsequent reheating temperature
are lower than the PQ scale.
Otherwise, the axion does not appear during the inflationary era and no CDM isocurvature
perturbation arises.
We assume that the inflaton itself does not generate non-Gaussianity, 
and that only the axion has an isocurvature fluctuation.
The CDM isocurvature perturbation $S$ is given as
\begin{eqnarray}
	S\; \simeq \; r \left[ \frac{2a_i \delta a}{a_*^2}+{\left( \frac{\delta a}{a_*} \right)}^2 \right], 
	\label{SCDM}
\end{eqnarray}
where $a_i=F_a\theta$ denotes the classical deviation from the potential minimum.
Here the second term in Eq.~(\ref{SCDM}) corresponds to non-Gaussian fluctuation.
Notice that here we have assumed that the axion sits near the potential minimum and the potential is
well approximated by the quadratic one.
While this is a reasonable assumption as will become clear, 
it is still possible that the axion starts oscillation from near the top of the potential.
In this case the prediction of non-Gaussianity will change~\cite{Kawasaki:2008mc}.
From Eq.~(\ref{fNLsimple}),
the non-linearity parameter $f_{\rm NL}^{(\rm iso)}$ is given by 
\begin{equation}
	\frac{6}{5}f_{\rm NL}^{(\rm iso)}=\frac{1}{27N_\phi^4} \left ( \frac{2r}{a_*^2} \right )^3 \left[ 
		 a_i^2 + \Delta_{\delta a}^2\ln (k_bL) \right].  \label{axionfNL}
\end{equation}
The coefficient of the trispectrum, $\tau_{\rm NL}^{(\rm iso)}$, is also calculated from Eq.~(\ref{tauNLsimple}) as
\begin{equation}
	\tau_{\rm NL}^{(\rm iso)}=\frac{1}{81N_\phi^6} \left ( \frac{2r}{a_*^2} \right )^4 \left[ 
		 a_i^2 + \Delta_{\delta a}^2\ln (k_tL) \right].  \label{axiontNL}
\end{equation}
Figs.~\ref{fig:NG-Fa10} and \ref{fig:NG-Fa12} show contours of 
$f_{\rm NL}^{(\rm iso)}=1$ and $100$ for $F_a = 10^{10}~$GeV (black solid lines)
and $\tau_{\rm NL}^{(\rm iso)}=10^3,10^5$ and $10^7$ (red dashed lines)
on $H_{\rm inf}$-$\theta$ plane, for $F_a=10^{10}$~GeV and $F_a=10^{12}$~GeV.
The gray shaded region is excluded from the isocurvature constraint.
In the blue region the PQ symmetry may be restored during inflation, and so,
neither isocurvature fluctuation nor non-Gaussianity will arise.
When the quantum fluctuation overcomes the classical deviation $(a_i^2 \ll \Delta_{\delta a}^2)$,
the second term in Eq.~(\ref{axionfNL}) and (\ref{axiontNL}) dominates.
In this limit, we find a consistency relation
\begin{equation}
	\tau_{\rm NL}^{(\rm iso)} = \left[ \frac{6}{5}f_{\rm NL}^{(\rm iso)} \right]^{4/3} \Delta_\zeta^{-2/3} C
	\sim 10^{3} \left[f_{\rm NL}^{(\rm iso)}\right ]^{4/3} C,
\end{equation}
where $C \equiv \ln (k_tL)/[\ln (k_bL)]^{4/3}$ is constant of order unity.
This relation is same as what is found in Ref.~\cite{Suyama:2008nt} for the ``ungaussiton'' scenario.
It is not surprising the same relation holds when the non-Gaussianity comes from isocurvature perturbation
if the quantum fluctuation dominates,
once we look at the similarity of the expression for $f_{\rm NL}$ and $\tau_{\rm NL}$ between 
the pure adiabatic and isocurvature cases.
As stated before, however,
it should be noted that their effects on CMB anisotropy are completely different.
In particular, non-Gaussianity of the isocurvature type at large scales
is expected to be significantly enhanced,
and hence these scenarios can be in principle distinguished.


\section{Conclusions} \label{conclusion}

The present observations show that the density perturbation of the Universe is
nearly adiabatic, but small admixture of the isocurvature fluctuation is still allowed.
Since the CDM (baryon) isocurvature perturbation contains information on how
it is produced,
the detection of the isocurvature perturbation and its non-Gaussianity will certainly give us a clue to
the origin of the density perturbation and underlying physics beyond the standard model.
In this paper we have derived the formulae for the bispectrum and trispectrum 
of the isocurvature perturbations.
Our formulae can be applied to the case where the adiabatic and isocurvature perturbations are
correlated.
As an explicit example, we have shown that the QCD axion can have large
isocurvature type non-Gaussianity. 
Therefore, the future CMB surveys may be able to detect the trispectrum and the bispectrum of the
isocurvature perturbation.
Such an analysis was performed in Ref.~\cite{Kogo:2006kh} for the adiabatic case
and it was found that the Planck survey will have a sensitivity $\tau_{\rm NL}^{\rm (adi)}\gtrsim 560$
at $2\sigma$ level.
This cannot be directly translated into the case of isocurvature trispectrum, since the effect on 
the CMB anisotropy is quite different between the adiabatic and isocurvature case.
We will pursue this issue as future work.

\section*{A.~Appendix} \label{app}

In this appendix we collect the form of bispectrum and trispectrum.

\subsection*{Bispectrum}

The each term in Eq.~(\ref{Bzeta}) is given by
\begin{equation}
\begin{split}
	B_{\zeta \zeta \zeta}& (k_1,k_2,k_3)  =
	N_aN_bN_{ab}\left[ P_{\delta \phi}(k_1) P_{\delta \phi}(k_2)+2~{\rm perms.} \right]\\
	&+N_{ab}N_{bc}N_{ca}\int \frac{d^3\vec k'}{(2\pi)^3}P_{\delta \phi}(k')
	P_{\delta \phi}(|\vec k_1+\vec k'|)P_{\delta \phi}(|\vec k_2-\vec k'|)\\
	&+\frac{N_{a}N_{bc}N_{abc}}{2}\left[ \int \frac{d^3\vec k'}{(2\pi)^3}P_{\delta \phi}(k')
	P_{\delta \phi}(|\vec k_1-\vec k'|)P_{\delta \phi}(k_2)
	+5~{\rm perms} \right],
\end{split}
\end{equation}
\begin{equation}
\begin{split}
	B_{\zeta \zeta S}& (k_1,k_2,k_3)  =
	(N_aN_bS_{ab}+2S_aN_bN_{ab})
	\left[ P_{\delta \phi}(k_1) P_{\delta \phi}(k_2)+2~{\rm perms.} \right]\\
	&+3N_{ab}N_{bc}S_{ca}\int \frac{d^3\vec k'}{(2\pi)^3}P_{\delta \phi}(k')
	P_{\delta \phi}(|\vec k_1+\vec k'|)P_{\delta \phi}(|\vec k_2-\vec k'|)\\
	&+\frac{1}{2}(S_{a}N_{bc}N_{abc}+N_{a}S_{bc}N_{abc}+N_{a}N_{bc}S_{abc} ) \\
	&~~~~~\times \left[ \int \frac{d^3\vec k'}{(2\pi)^3}P_{\delta \phi}(k')
	P_{\delta \phi}(|\vec k_1-\vec k'|)P_{\delta \phi}(k_2)
	+5~{\rm perms} \right],
\end{split}
\end{equation}
\begin{equation}
\begin{split}
	B_{\zeta SS}& (k_1,k_2,k_3)  =
	(S_aS_bN_{ab}+2N_aS_bS_{ab})
	\left[ P_{\delta \phi}(k_1) P_{\delta \phi}(k_2)+2~{\rm perms.} \right]\\
	&+3N_{ab}S_{bc}S_{ca}\int \frac{d^3\vec k'}{(2\pi)^3}P_{\delta \phi}(k')
	P_{\delta \phi}(|\vec k_1+\vec k'|)P_{\delta \phi}(|\vec k_2-\vec k'|)\\
	&+\frac{1}{2}(S_{a}S_{bc}N_{abc}+S_{a}N_{bc}S_{abc}+N_{a}S_{bc}S_{abc}) \\
	&~~~~~\times \left[ \int \frac{d^3\vec k'}{(2\pi)^3}P_{\delta \phi}(k')
	P_{\delta \phi}(|\vec k_1-\vec k'|)P_{\delta \phi}(k_2)
	+5~{\rm perms} \right],
\end{split}
\end{equation}
\begin{equation}
\begin{split}
	B_{SSS}& (k_1,k_2,k_3)  =
	S_aS_bS_{ab}\left[ P_{\delta \phi}(k_1) P_{\delta \phi}(k_2)+2~{\rm perms.} \right]\\
	&+S_{ab}S_{bc}S_{ca}\int \frac{d^3\vec k'}{(2\pi)^3}P_{\delta \phi}(k')
	P_{\delta \phi}(|\vec k_1+\vec k'|)P_{\delta \phi}(|\vec k_2-\vec k'|)\\
	&+\frac{S_{a}S_{bc}S_{abc}}{2}\left[ \int \frac{d^3\vec k'}{(2\pi)^3}P_{\delta \phi}(k')
	P_{\delta \phi}(|\vec k_1-\vec k'|)P_{\delta \phi}(k_2)
	+5~{\rm perms} \right].
\end{split}
\end{equation}

\subsection*{Trispectrum}

The each term in Eq.~(\ref{Tzeta}) is given by
\begin{equation}
\begin{split}
	T_{\zeta \zeta \zeta \zeta}& (k_1,k_2,k_3,k_4)  =
	N_aN_bN_{ac}N_{bc}
	\left[ P_{\delta \phi}(k_1) P_{\delta \phi}(k_2)P_{\delta \phi}(k_{13})+11~{\rm perms.} \right] \\
	&+N_{a}N_{b}N_{c}N_{abc} \left[ P_{\delta \phi}(k_1)
	P_{\delta \phi}(k_2)P_{\delta \phi}(k_3) + 3~{\rm perms.}\right] \\
	&+N_{ab}N_{bc}N_{cd}N_{da}
	\left[ \int \frac{d^3\vec p}{(2\pi)^3}P_{\delta \phi}(p)
	P_{\delta \phi}(|\vec k_1-\vec p|)P_{\delta \phi}(|\vec k_2+\vec p|)
	P_{\delta \phi}(|\vec p-\vec k_1-\vec k_3|) +2~{\rm perms.} \right]  \\
	&+\frac{N_{abc}N_{abd}N_{c}N_{d}}{2}
	\left[ \int \frac{d^3\vec p}{(2\pi)^3}P_{\delta \phi}(p)
	P_{\delta \phi}(|\vec k_1+\vec k_3-\vec p|)P_{\delta \phi}(k_1)
	P_{\delta \phi}(k_2) +11~{\rm perms.} \right]  \\
	&+\frac{N_{abcd}N_{ab}N_{c}N_{d}}{2}
	\left[ \int \frac{d^3\vec p}{(2\pi)^3}P_{\delta \phi}(p)
	P_{\delta \phi}(|\vec k_2+\vec p|)P_{\delta \phi}(k_3)
	P_{\delta \phi}(k_4) +11~{\rm perms.} \right]  \\
	&+N_{abc}N_{ab}N_{cd}N_{d}
	\left[ \int \frac{d^3\vec p}{(2\pi)^3}P_{\delta \phi}(p)
	P_{\delta \phi}(|\vec k_2+\vec p|)P_{\delta \phi}(|\vec k_1+\vec k_2|)
	P_{\delta \phi}(k_4) +11~{\rm perms.} \right]  \\
	&+N_{abc}N_{ad}N_{cd}N_{b}
	\left[ \int \frac{d^3\vec p}{(2\pi)^3}P_{\delta \phi}(p)
	P_{\delta \phi}(|\vec k_2+\vec p|)P_{\delta \phi}(|\vec k_1+\vec k_4-\vec p|)
	P_{\delta \phi}(k_4) +11~{\rm perms.} \right]  \\
\end{split}
\end{equation}
\begin{equation}
\begin{split}
	T_{\zeta \zeta \zeta S}& (k_1,k_2,k_3,k_4)  =
	2(N_aS_bN_{ac}N_{bc} + N_aN_bN_{ac}S_{bc})
	\left[ P_{\delta \phi}(k_1) P_{\delta \phi}(k_2)P_{\delta \phi}(k_{13})+11~{\rm perms.} \right] \\
	&+(3N_{a}N_{b}S_{c}N_{abc}+S_{abc}N_aN_bN_c) \left[ P_{\delta \phi}(k_1)
	P_{\delta \phi}(k_2)P_{\delta \phi}(k_3) + 3~{\rm perms.}\right] \\
	&+4N_{ab}N_{bc}N_{cd}S_{da} \\
	&~~~~~\times \left[ \int \frac{d^3\vec p}{(2\pi)^3}P_{\delta \phi}(p)
	P_{\delta \phi}(|\vec k_1-\vec p|)P_{\delta \phi}(|\vec k_2+\vec p|)
	P_{\delta \phi}(|\vec p-\vec k_1-\vec k_3|) +2~{\rm perms.} \right]  \\
	&+\frac{1}{2}(2N_{abc}S_{abd}N_{c}N_{d}+2N_{abc}N_{abd}N_{c}S_{d}  )\\
	&~~~~~\times \left[ \int \frac{d^3\vec p}{(2\pi)^3}P_{\delta \phi}(p)
	P_{\delta \phi}(|\vec k_1+\vec k_3-\vec p|)P_{\delta \phi}(k_1)
	P_{\delta \phi}(k_2) +11~{\rm perms.} \right]  \\
	&+\frac{1}{2}(S_{abcd}N_{ab}N_{c}N_{d}+N_{abcd}S_{ab}N_{c}N_{d}
	   +2N_{abcd}N_{ab}N_{c}S_{d})
	\\ &~~~~~\times \left[ \int \frac{d^3\vec p}{(2\pi)^3}P_{\delta \phi}(p)
	P_{\delta \phi}(|\vec k_2+\vec p|)P_{\delta \phi}(k_3)
	P_{\delta \phi}(k_4) +11~{\rm perms.} \right]  \\
	&+(S_{abc}N_{ab}N_{cd}N_{d}+N_{abc}S_{ab}N_{cd}N_{d}
		+N_{abc}N_{ab}S_{cd}N_{d}+N_{abc}N_{ab}N_{cd}S_{d}  )\\
	&~~~~~\times \left[ \int \frac{d^3\vec p}{(2\pi)^3}P_{\delta \phi}(p)
	P_{\delta \phi}(|\vec k_2+\vec p|)P_{\delta \phi}(|\vec k_1+\vec k_2|)
	P_{\delta \phi}(k_4) +11~{\rm perms.} \right]  \\
	&+(S_{abc}N_{ad}N_{cd}N_{b}+2N_{abc}S_{ad}N_{cd}N_{b}+N_{abc}N_{ad}N_{cd}S_{b}) \\
	&~~~~~\times \left[ \int \frac{d^3\vec p}{(2\pi)^3}P_{\delta \phi}(p)
	P_{\delta \phi}(|\vec k_2+\vec p|)P_{\delta \phi}(|\vec k_1+\vec k_4-\vec p|)
	P_{\delta \phi}(k_4) +11~{\rm perms.} \right]  \\
\end{split}
\end{equation}
\begin{equation}
\begin{split}
	T_{\zeta \zeta SS}& (k_1,k_2,k_3,k_4)  =
	(N_aN_bS_{ac}S_{bc}+2N_aN_{ab}S_{bc}S_c+2N_aS_{ab}N_{bc}S_c+S_aS_bN_{ac}N_{bc}) \\
	&~~~~~\times \left[ P_{\delta \phi}(k_1) P_{\delta \phi}(k_2)
	P_{\delta \phi}(k_{13})+11~{\rm perms.} \right] \\
	&+(3N_{a}S_{b}S_{c}N_{abc}+3S_{abc}S_aN_bN_c) \left[ P_{\delta \phi}(k_1)
	P_{\delta \phi}(k_2)P_{\delta \phi}(k_3) + 3~{\rm perms.}\right] \\
	&+(4N_{ab}N_{bc}S_{cd}S_{da} +2N_{ab}S_{bc}N_{cd}S_{da}  )\\
	&~~~~~\times \left[ \int \frac{d^3\vec p}{(2\pi)^3}P_{\delta \phi}(p)
	P_{\delta \phi}(|\vec k_1-\vec p|)P_{\delta \phi}(|\vec k_2+\vec p|)
	P_{\delta \phi}(|\vec p-\vec k_1-\vec k_3|) +2~{\rm perms.} \right]  \\
	&+\frac{1}{2}(S_{abc}S_{abd}N_{c}N_{d}+ 2N_{abc}S_{abd}N_{c}S_{d}
		+2N_{abc}S_{abd}S_{c}N_{d}+N_{abc}N_{abd}S_{c}S_{d}  ) \\
	&~~~~~\times \left[ \int \frac{d^3\vec p}{(2\pi)^3}P_{\delta \phi}(p)
	P_{\delta \phi}(|\vec k_1+\vec k_3-\vec p|)P_{\delta \phi}(k_1)
	P_{\delta \phi}(k_2) +11~{\rm perms.} \right]  \\
	&+\frac{1}{2}(N_{abcd}N_{ab}S_{c}S_{d}+2N_{abcd}S_{ab}N_{c}S_{d}
	      +2S_{abcd}N_{ab}N_{c}S_{d}+S_{abcd}S_{ab}N_{c}N_{d})
	\\ &~~~~~\times \left[ \int \frac{d^3\vec p}{(2\pi)^3}P_{\delta \phi}(p)
	P_{\delta \phi}(|\vec k_2+\vec p|)P_{\delta \phi}(k_3)
	P_{\delta \phi}(k_4) +11~{\rm perms.} \right]  \\
	&+(S_{abc}S_{ab}N_{cd}N_{d}+S_{abc}N_{ab}N_{cd}S_{d}
		+S_{abc}N_{ab}S_{cd}N_{d}+N_{abc}S_{ab}N_{cd}S_{d} \\
	&~~~~~~~~~~+N_{abc}S_{ab}S_{cd}N_{d}+N_{abc}N_{ab}S_{cd}S_{d}  )\\
	&~~~~~\times \left[ \int \frac{d^3\vec p}{(2\pi)^3}P_{\delta \phi}(p)
	P_{\delta \phi}(|\vec k_2+\vec p|)P_{\delta \phi}(|\vec k_1+\vec k_2|)
	P_{\delta \phi}(k_4) +11~{\rm perms.} \right]  \\
	&+(2N_{abc}N_{ad}S_{cd}S_{b}+N_{abc}S_{ad}S_{cd}N_{b}
		+S_{abc}N_{ad}N_{cd}S_{b}+2S_{abc}N_{ab}S_{cd}N_{d}) \\
	&~~~~~\times \left[ \int \frac{d^3\vec p}{(2\pi)^3}P_{\delta \phi}(p)
	P_{\delta \phi}(|\vec k_2+\vec p|)P_{\delta \phi}(|\vec k_1+\vec k_4-\vec p|)
	P_{\delta \phi}(k_4) +11~{\rm perms.} \right],  
\end{split}
\end{equation}
and $T_{\zeta SSS}$ and $T_{SSSS}$ are obtained by interchanging $N$ and $S$ in 
the expression of $T_{\zeta \zeta \zeta S}$ and $T_{\zeta\zeta\zeta\zeta}$, respectively.

\section*{Acknowledgment}

K.N. would like to thank the Japan Society for the Promotion of
Science (JSPS) for financial support.
The work of F.T. was supported by JSPS (21740160).
This work is supported by Grant-in-Aid for Scientific research from the Ministry of Education,
Science, Sports, and Culture (MEXT), Japan, No.14102004 (M.K.)
and also by World Premier International
Research Center Initiative (WPI Initiative), MEXT, Japan.



{}

\end{document}